\documentclass[12pt]{article}
\begin{document}
\begin{center}
{\large \bf SEMI-PHENOMENOLOGICAL APPROACH TO THE ESTIMATE OF CP EFFECTS
IN $K^{\pm} \to 3 \pi $ DECAYS}\\
\vspace{5mm} Evgeny Shabalin \footnote{e-mail:  shabalin@heron.itep.ru}\\
Institute for Theoretical and Experimental Physics, Moscow, Russia
\end{center} \vspace{3cm} \begin{abstract} The amplitudes of the $K^{\pm}
\to 3\pi$ and $K \to 2\pi$ decays are expressed in terms of different
combinations of one and the same set  of CP-conserving and CP-odd
parameters. Extracting the magnitudes of these parameters from the data on
$K \to 2\pi$ dacays, we estimate an expected CP-odd difference between the
values of the slope parameters $g^+$ and $g^-$ of the energy distributions
of "odd" pions in $K^+ \to \pi^+\pi^+ \pi^-$  and $K^- \to \pi^-\pi^-
\pi^+$ decays.  \end{abstract} \newpage \section{Introduction} The
observation of CP effects in $K^{\pm} \to 3\pi$ decays would allow to
understand better how the mechanisms of CP violation work.

Now the Collaboration NA48/2 is ready to begin a search for such effect
with accuracy $\delta(\frac{g^+ -g^-}{g^+ +g^-})\le 2\cdot 10^{-4}$.

Contrary to the case of $K_L \to 2\pi$ decay where CP violates both in
$\Delta S =2$ and $\Delta S=1$ transitions, in the $K^{\pm} \to 3\pi$
decays, only the last (so-called "direct" ) CP violation takes place.
Experimentally, an existance of the direct CP violation in $ K_L \to 2\pi$
decays, predicted by Standard Model (SM) and characterised by the
parameter $\varepsilon'$ is establised: $\varepsilon'/\varepsilon =(1.66
\pm 0.16)10^{-3}$. But the large uncertainties in the theoretical
predictions
$$
\frac{\varepsilon'}{\varepsilon}=(17^{+14}_{-10})10^{-4} \quad \cite{1},
\qquad  \frac{\varepsilon'}{\varepsilon}=(1.5-31.6)10^{-4} \quad
\cite{2} $$ do not allow to affirm that the contributions from the sources
of CP violation beyond the Kobayashi-Maskawa phase are excluded.

 To avoid the uncertainties in the theoretical calculation of the
ingredients of the theory, we use the following procedure. We
express the amplitudes of $K_L \to 2\pi$ and $K^{\pm} \to 3\pi$ in terms
of one and the same set of parameters, and calculating $g^+ -g^-$ we
use the magnitudes of these parameters extracted from data on $K \to
2\pi$ decays.
\section{The scheme of calculation}
A theory of $\Delta S=1$ non-leptonic decays is based on the effective
lagrangian \cite{3}
\begin{equation}
L(\Delta S=1) =\sqrt{2}G_F\sin\theta_C \cos \theta_C \sum c_iO_i
\label{1}
\end{equation}
where
\begin{equation}
O_1=\bar s_L \gamma_{\mu}d_L \cdot \bar u_L \gamma_{\mu} u_L - \bar s_L
\gamma_{\mu} u_L \cdot \bar u_L \gamma_{\mu} d_L \qquad(\{8_f\}, \Delta I =
1/2)
\end{equation}
\begin{eqnarray}
O_2=\bar s_L \gamma_{\mu} d_L \cdot \bar u_L \gamma_{\mu} u_L + \bar s_L
\gamma_{\mu} u_L \cdot \bar u_L \gamma_{\mu} d_L +2\bar s_L \gamma_{\mu}
d_L \cdot \bar d_L \gamma_{\mu} d_L \nonumber  \\
\qquad {}+2 \bar s_L \gamma_{\mu} d_L \cdot \bar s_L \gamma_{\mu} s_L
\qquad (\{8_d\}, \Delta I =1/2)
\end{eqnarray}
\begin{eqnarray}
O_3=\bar s_L \gamma_{\mu} d_L \cdot \bar u \gamma_{\mu} u_L +\bar s
\gamma_{\mu} u_L \cdot \bar u \gamma_{\mu} d_L + 2 \bar s_L \gamma{\mu}
d_L \cdot \bar d_L \gamma_{\mu} d_L \nonumber \\
\qquad {}-3 \bar s_L \gamma_{\mu} d_L \cdot \bar s_L \gamma_{\mu} s_L
\qquad (\{27\}, \Delta I=1/2)
\end{eqnarray}
\begin{eqnarray}
O_4=\bar s_L \gamma_{\mu} d_L \cdot \bar u \gamma_{\mu} u_L + \bar s_L
\gamma_{\mu} u_L \cdot \bar u_L \gamma_{\mu} d_L - \nonumber \\
\qquad {}-\bar s_L \gamma_{\mu} d_L \cdot \bar d_L \gamma_{\mu} d_L
\qquad (\{27\}, \Delta I=3/2)
\end{eqnarray}
\begin{equation}
O_5= \bar s_L \gamma_{\mu} \lambda^a d_L (\sum_{q=u,d,s} \bar q_R
\gamma_{\mu} \lambda^a q_R) \qquad (\{8\},  \Delta I=1/2)
\end{equation}
\begin{equation}
O_6= \bar s_L \gamma_{\mu} d_L (\sum_{q=u,d,s} \bar q_R \gamma_{\mu} q_R)
\qquad (\{8 \}, \Delta I = 1/2)
\end{equation}
This set is sufficient for calculation of the CP-even parts of the
amplitudes under consideration. To calculate the CP-odd parts , it is
necessary to add the so-called electroweak contributions originated by the
operators $O_7, O_8$:
\begin{equation}
O_7=\frac{3}{2} \bar s\gamma_{\mu}(1+\gamma_5)d\cdot (\sum_{q=u,d,s}e_q
\bar q \gamma_{\mu}(1-\gamma_5)q) \qquad  (\Delta I=1/2, 3/2)
\end{equation} \begin{equation} O_8=-12\sum_{q=u,d,s} e_q (\bar s_L
q_R)(\bar q_R d_L),\quad e_q=(\frac{2}{3}, -\frac{1}{3}, -\frac{1}{3})
,\quad (\Delta I=1/2, 3/2) \end{equation} The coefficients $ c_{5-8}$ have
the imaginary parts necessary for CP violation.

The bosonization of these operators can be done using the relations
\cite{4}
\begin{equation}
\bar
q_j(1+\gamma_5)q_k=-\frac{1}{\sqrt{2}}F_{\pi}r\left(U-\frac{1}{\Lambda^2}
\partial^2 U \right)_{kj}
\end{equation}
\begin{equation}
\bar q_j \gamma_{\mu}(1+\gamma_5) q_k=i[\left(\partial_{\mu}U
\right)U^{\dag} - U \left( \partial_{\mu} U^{\dag}  \right)-
\frac{rF_{\pi}}{\sqrt{2}\Lambda^2} \left(m(\partial_{\mu} U^{\dag}
)-( \partial_{\mu} U)m \right)]_{kj}
\end{equation}
if the non-linear realization of chiral symmetry is used:
\begin{equation}
U=\frac{F_{\pi}}{\sqrt{2}}\left(1+\frac{i\sqrt{2} \hat
\pi}{F_{\pi}}-\frac{\hat \pi^2}{F_{\pi}^2} +a_3\left(\frac{i\hat
\pi}{\sqrt{2}F_{\pi}}\right)^3 +2(a_3-1)\left(\frac{i\hat
\pi}{\sqrt{2}F_{\pi}}\right)^4 +....\right)
\end{equation}
where
\begin{equation}
\hat \pi=
\left(
\begin{array}{lll}
\frac{\pi_0}{\sqrt{3}}+\frac{\pi_8}{\sqrt{6}}+\frac{\pi_3}{\sqrt{2}} &
\quad \pi^+ & \quad K^+ \\
\pi^- & \quad
\frac{\pi_0}{\sqrt{3}}+\frac{\pi_8}{\sqrt{6}}-\frac{\pi_3}{\sqrt{2}} &
\quad K^0 \\ K^- & \quad \bar K^0 & \quad
\frac{\pi_0}{\sqrt{3}}-\frac{2\pi_8}{\sqrt{6}} \end{array} \right)
\end{equation}
The PCAC condition demands $a_3=0$  \cite{5} and we adopt this condition,
bearing in mind that, on mass shell, the values of the mesonic amplitudes
are independent of $a_3$.

Using also the relations between matrices in the colour space
$$
\begin{array}{lll}
\delta^{\alpha}_{\beta} \delta^{\gamma}_{\delta} =\frac{1}{3}
\delta^{\alpha}_{\delta}
\delta^{\gamma}_{\beta}+\frac{1}{2}\lambda^{\alpha}_{\delta}
\lambda^{\gamma}_{\beta} \\
\lambda^{\alpha}_{\beta}
\lambda^{\gamma}_{\delta}=\frac{16}{9}\delta^{\alpha}_{\beta}
\delta^{\gamma}_{\beta} -\frac{1}{3} \lambda^{\alpha}_{\delta}
\lambda^{\gamma}_{\beta}
\end{array}
$$
and the Fierz transformation relation
$$
\bar s \gamma_{\mu}(1+\gamma_5) d \cdot \bar q \gamma_{\mu}(1-\gamma_5)q=
-2\bar s(1-\gamma_5)q \cdot \bar q(1+\gamma_5)d
$$
and representing $M(K \to 2\pi)$ in the form
\begin{eqnarray}
M(K^0_1 \to \pi^+\pi^-)=A_0e^{i\delta_0}-A_2e^{i\delta_2} \\
M(K^0_1 \to \pi^0\pi^0)=A_0e^{i\delta_0}+2A_2e^{i\delta_2} \\
M(K^+ \to \pi^+\pi^0) =-\frac{3}{2} A_2 e^{i\delta_2}
\end{eqnarray}
we obtain
\begin{eqnarray}
A_0=G_FF_{\pi}\sin \theta_C \cos\theta_C\frac{m^2_K-m^2_{\pi}}{\sqrt{2}}
\cdot[c_1-c_2-c_3 +\frac{32}{9}\beta(Re \tilde c_5 +i Im \tilde c_5)] \\
A_2=G_FF_{\pi}\sin \theta_C \cos \theta_C \frac{m^2_K-m^2_{\pi}}{\sqrt{2}}
\cdot [c_4 +i \frac{2}{3}\beta \Lambda^2 Im\tilde
c_7(m^2_K-m^2_{\pi})^{-1}]
\end{eqnarray}
where
$$
\tilde c_5=c_5+\frac{3}{16} c_6;\quad \tilde c_7=c_7+3c_8; \quad
\beta=\frac{2m^4_{\pi}}{\Lambda^2 (m_u+m_d)^2}.
$$
The contributions from $\tilde c_7 O_7$ into $ReA_0$ and $ImA_0$ are small
because $\tilde c_7/\tilde c_5 \sim \alpha_{em}$ and we neglected these
corrections. From data on widths of $K \to 2\pi$ decays we obtain
\begin{equation}
c_4 =0.328;\quad c_1-c_3-c_3+\frac{32}{9} \beta Re \tilde c_5 =-10.13.
\end{equation}
At $c_1-c_2-c_3=-2.89$ \cite{3}, \cite{6} and $\beta=6.68$ we obtain
\begin{equation}
\tilde c_5=-0.305.
\end{equation}
From the expression for $A_2$, it is seen that the contribution
of the operators $O_{7,8}$  is enlarged by the factor $\Lambda^2/m^2_K$ in
comparison with the rest operators contribution.

Using the general relation
\begin{equation}
\varepsilon'=ie^{i(\delta_2-\delta_0)}\left[-\frac{ImA_0}{ReA_0}+\frac{Im
A_2}{Re A_2}\right]\cdot |\frac{A_2}{A_0}|
\end{equation}
and the experimental value $\varepsilon'=(3.4 \pm 0.45)10^{-6}$ we come to
the relation
\begin{equation}
-\frac{Im \tilde c_5}{Re \tilde c_5} \left(1-\Omega_{\eta,\eta'} +20.66
\frac{Im \tilde c_7}{Im \tilde c_5} \right)=1.48 \cdot 10^{-4}.
\end{equation}
where $\Omega_{\eta,\eta'}$ takes into account the effects of $K^0
\to\pi^0 \eta (\eta') \to \pi^0\pi^0 $ transitions.

The naive estimate gives
\begin{equation}
-\frac{Im \tilde c_5}{Re \tilde c_5} \approx 1.7 s_2 s_3 \sin \delta
\end{equation}
where $s_2, s_3$ and $\delta$ are the parameters of CKM matrix.
At
$$
4.6 \cdot 10^{-4} \le s_2 s_3 \le 6.7 \cdot 10^{-4} \qquad \mbox
{(Landsberg'2002)}
$$
\begin{equation}
\frac{Im \tilde c_5}{Re \tilde c_5}=(-9.6 \pm 1.8)10^{-4}\sin \delta
\end{equation}
and
\begin{equation}
\frac{Im \tilde c_7}{Im \tilde c_5}=
\left \{
\begin{array}{ll}
-0.026 \quad \mbox{for} \quad \Omega_{\eta,\eta'}=0.3 \\
-0.041 \quad \mbox{for} \quad \Omega_{\eta,\eta'}=0
\end{array}
\right.
\end{equation}
\section{Decay $K^{\pm} \to \pi^{\pm} \pi^{\pm} \pi^{\mp}$}
In the leading $p^2$ approximation
\begin{equation}
M(K^+ \to \pi^+(p_1) \pi^+(p_2) \pi^-(p_3))=k[1+ia_{KM} +\frac{1}{2} g
Y(1+ib_{KM})+....],
\end{equation}
where
\begin{equation}
k=G_F \sin \theta_C \cos \theta_C m^2_K c_0 (3\sqrt{2})^{-1}
\end{equation}
\begin{equation}
a_{KM}=\left[\frac{32}{9} \beta Im \tilde c_5 + 4\beta Im \tilde c_7
\left(\frac{3\Lambda^2}{2m^2_K}+2 \right) \right]/c_0
\end{equation}
\begin{equation}
b_{KM}=\left[\frac{32}{9} \beta Im \tilde c_5 +8\beta Im \tilde c_7
\right]/(c_0+9c_4)
\end{equation}
\begin{equation}
g=-\frac{3m^2_{\pi}}{2m^2_K}(1+9c_4/c_0) ,\qquad Y=(s_3-s_0)/m^2_{\pi}
\end{equation}
\begin{equation}
c_0=c_1-c_2-c_3-c_4 +\frac{32}{9} \beta Re \tilde c_5 =-10.46
\end{equation}

As the field $K^+$ is the complex one and its phase is arbitrary, we can
replace $K^+$ by $K^+(1+ia_{KM})(\sqrt{1+a^2_{KM}})^{-1}$. Then
\begin{equation}
M(K^+ \to \pi^+ \pi^+ \pi^-(p_3))
=k[1+\frac{1}{2}gY(1+i(b_{KM}-a_{KM}))+...]
\end{equation}
Though this expression contains the imaginary CP-odd part, it does not
lead to observable CP effects.  Such effects arise due to interference
between CP-odd imaginary part with the CP-even imaginary part produced by
rescattering of the final pions.  Then
\begin{equation}
M(K^+ \to \pi^+ \pi^+ \pi^-) = k[1+ia +\frac{1}{2}
gY(1+ib+i(b_{KM}-a_{KM}) +...]
\end{equation}
where $a$ and $b$ are corresponding CP-even imaginary parts of the
amplitude. These parts can be estimated ( in $p^2$)
approximation calculating the imaginary part of the two-pion loop diagrams
with $$ \begin{array}{lll} M(\pi^+(r_2) \pi^-(r_3) \to \pi^+(p_2)
\pi^-(p_3)) =F_{\pi}^{-2}[(p_2+p_3)^2 + (r_2-p_2)^2 -2m^2_{\pi}] \\
M(\pi^0(r_2) \pi^0(r_3) \to \pi^+(p_2) \pi^- (p_3))=F^{-2}_{\pi}
[(p_2+p_3)^2 -m^2_{\pi}] \\ M(\pi^+(r_1) \pi^+(r_2) \to \pi^+(p_1)
\pi^+(p_2))=F^{-2}_{\pi}[(r_1-p_1)^2 +(r_1-p_2)^2 -2m^2_{\pi}] \end{array}
$$
Then we find:
\begin{equation}
a=  0.12065; \qquad b=0.714
\end{equation}
Using the definition of the slope paramater
\begin{equation}
|M(K^{\pm} \to \pi^{\pm} \pi^{\pm} \pi^{\mp}(p_3))|^2 \sim
1+\frac{g}{1+a^2}Y\left(1+ab \pm a(b_{KM}-a_{KM})\right)+...
\end{equation}
we find
\begin{equation}
R_g \equiv \frac{g^+-g^-}{g^+ +g^-}= \frac{a(b_{KM}-a_{KM})}{1+ab}
\end{equation}
At the fixed above numerical values of the parameters we obtain
\begin{equation}
\left(R_g \right) _{p^2} =0.030 \frac{Im \tilde
c_5}{Re \tilde c_5}(1-14.9 \frac{Im \tilde c_7}{Im \tilde c_5}) =-(4 \pm
0.75 )\cdot 10^{-5} \sin \delta \end{equation}
This numerical result is obtained for $\Omega_{\eta,\eta'}=0.3$. For zero
magnitude of this parameter, a value would be $(-4.6 \pm 0.86)10^{-5}$.
\section{The role of $p^4$ and other corrections}
The corrections to the result obtained in the conventional chiral theory
up to leading $p^2$ approximation are of two kinds. The first kind
corrections are connected with a necessity to take into account the
observed enlargement of $S$-wave $I=0$ $\pi \pi$ amplitude having no
explanation in conventional chiral theory.
The corrections of
the second kind are the $p^4$ corrections. As it was argued in \cite{7},
\cite{8} both kinds corrections can be properly estimated in the framework
of special linear $U(3)_L \otimes U(3)_R$ $\sigma$ model with broken
chiral symmetry. The above mentioned enlargement of $S$ wave in this model
is originated by mixing between the $\bar q q$ and $(G^a_{\mu\nu})^2$
states. In such a model
$$
U=\hat \sigma + i \hat \pi
$$
where $\hat \sigma$ is $3\times 3$ matrix of scalar partners of the mesons
of pseudoscalar nonet. The relations between diquark combinations and
spinless fields are as given by eqs.(10) ,(11), but without  the terms
proportional to $\Lambda^{-2}$. Such contributions in $\sigma$ model
appear from an expansion of the intermediate scalar mesons propagators.
The parameter $\Lambda^2$ acquires a sence of difference
$m^2_{a_0(980)}-m^2_{\pi}$. The strength of mixing between the isosinglet
$\sigma$  meson and corresponding gluonic state is characterised
by the parameter $\xi$.

If the $p^2$ approximation gives \begin{equation}
(k)_{p^2}=1.495 \cdot 10^{-4}, \qquad (g)_{p^2} =-0.172 \end{equation}
instead of \begin{equation}(k)_{exp}=1.72 \cdot 10^{-4}, \qquad
(g)_{exp}=-0.2154 \pm 0.0035, \end {equation} the corrected values of these
CP-even parameters of $K^+ \to \pi^+ \pi^+ \pi-$ amplitude practically
coincide with the experimental ones \cite{7}:  \begin{equation}
(k)_{(p^2+p^4;\; \xi=-0.225)}=1.72 \cdot 10^{-4}, \qquad (g)_{(p^2+p^4;\;
\xi =-0.225)}=-0.21.  \end{equation} The meaning of the parameter $\xi$ is
explained in \cite{7},\cite{8}. The expressions for the corrected $\pi \pi
\to \pi \pi$ amplitudes are presented in \cite{8}.

Calculating the CP-even imaginary part of the $ K^{\pm} \to \pi^{\pm}
\pi^{\pm} \pi^{\mp}$ amplitude originated by two-pion intermediate states,
we obtain
\begin{equation}
a(p^2+p^4; \;\xi=-0.225)=0.16265
\end{equation}
\begin{equation}
b(p^2+p^4;\; \xi=-0.225) =0.762
\end{equation}

An estimate of the parameter $a$ can be obtained also without any
calculations using the circumstance that at $\sqrt{s}=\sqrt{s_0}$, the
only significant phase shift is $\delta^0_0$. The rest phase shifts are
very small:  $|\delta^2_0(s_0)|<1.8^{\circ}$ and $\delta^1_1(s_0)
<0.3^{\circ} $ \cite{9}.  Then, according to eq.(33), $a \approx \tan
\delta^0_0(s_0)$, or $a=0.13 \pm 0.05$, if $\delta^0_0(s_0)=(7.50 \pm
2.85)^{\circ}$ \cite{10} and $a=0.148 \pm 0.018$, if $\delta^0_0(s_0)=(8.4
\pm 1.0)^{\circ}$ \cite{11}. These results coincide inside the error bars
with the result (41).  The corrected magnitude of $R_g$ is
\begin{equation}
\left(R_g \right)_{(p^2+p^4; \; \xi=-0.225)}=0.039\frac{Im \tilde c_5}{Re
\tilde c_5} \left(1-11.95 \frac{Im \tilde c_7}{Im \tilde c_5} \right)\sin
\delta=- (4.9 \pm 0.9)10^{-5} \sin \delta.
\end{equation}
This result is by 22\% larger in absolute magnitude than that calculated
in the leading approximation.  Therefore, we come to conclusion that the
corrections to the result obtained in the framework of conventional chiral
theory to the leading approximation are not negligible (20-30 \%), but not
so large, as it was declared in \cite{12}.  \section{Conclusion} If the
arguments against the correctness of eq.(23) will not be found, an
expected value of $R_g$ in the Standard Model is not larger in absolute
magnitude than $6 \cdot 10^{-5} \sin \delta$.

\end{document}